\magnification=1200
\overfullrule=0mm
\centerline{\bf Matrices coupled in a chain. II. Spacing functions}
\vskip 3mm

\centerline{Gilbert Mahoux\footnote\S {e-mail address: 
mahoux@spht.saclay.cea.fr }, 
Madan Lal Mehta\footnote* {Member of Centre National de la Recherche 
Scientifique,  France} 
\footnote\dag {e-mail address: mehta@spht.saclay.cea.fr}
and
Jean-Marie Normand\footnote{**}{e-mail address: norjm@spht.saclay.cea.fr}}
\centerline{CEA/Saclay, Service de Physique Th\'eorique}
\centerline{F-91191 Gif-sur-Yvette Cedex, FRANCE}
\vskip 1cm
\noindent {\bf Abstract.} For the eigenvalues of $p$ complex hermitian 
$n\times n$ matrices coupled in a chain, we 
give a method of calculating the spacing functions. This is a 
generalization of the one matrix case which has been known for a long time.
\vskip 19mm

\noindent {\bf 1. Introduction}
\vskip 3mm

\noindent Let us recall here a few facts concerning the case of one matrix.
For a $n\times n$ complex hermitian matrix $A$ with matrix 
elements probability density 
${\rm exp}[-{\rm tr}\ V(A)]$, the probability density of its eigenvalues 
${\bf x}:= \{x_1, x_2, ..., x_n\}$ is [1] 
$$ \eqalignno{
F({\bf x}) & \propto {\rm exp}\left[-\sum_{j=1}^n V(x_j)\right] 
\prod_{1\le i<j\le n}(x_j-x_i)^2, & (1.1a) \cr 
& \propto \det[K(x_i,x_j)]_{i,j=1,...,n}, & (1.1b) \cr} $$
where $V(x)$ is a real polynomial of even order, the coefficient of the 
highest power being positive; $K(x,y)$ is defined by
$$K(x,y) := \exp \left[-{1\over 2}V(x)-{1\over 2}V(y)\right] \sum_{i=0}^{n-1} 
{1\over h_i} P_i(x) P_i(y), \eqno(1.2) $$
$P_i(x)$ is a real polynomial of degree $i$ and the polynomials are 
chosen orthogonal with the weight $\exp[-V(x)]$, 
$$ \int P_i(x) P_j(x) \exp[-V(x)] dx = h_i \delta_{ij}. \eqno (1.3) $$ 
Here and in what follows, all the integrals are taken from $-\infty$ 
to $+\infty$, unless explicitly stated otherwise.

The correlation function $R_k(x_1,...,x_k)$, i.e. the density of ordered 
sets of $k$ eigenvalues within small intervals around $x_1$, ..., $x_k$, 
ignoring the other eigenvalues, is 
$$ \eqalignno{
 R_k(x_1,...,x_k) & := {n!\over (n-k)!} \int F({\bf x}) dx_{k+1}...dx_n 
& \cr 
& = \det[K(x_i,x_j)]_{i,j=1,...,k}. & (1.4) \cr}$$

The spacing function $E(k,I)$, i.e. the 
probability that a chosen domain $I$ contains exactly $k$ eigenvalues 
$(0\le k\le n)$, is 
$$ \eqalignno{
E(k,I) & := {n!\over k!(n-k)!} \int F({\bf x})\left[\prod_{j=1}^k\chi(x_j)
\right] \left[\prod_{j=k+1}^n\left[1-\chi(x_j)\right]\right]  
dx_1...dx_n  & \cr
& =  {1\over k!}\left.\left({d\over dz}\right)^k R(z,I)
\right\vert_{z=-1}, & (1.5) \cr} $$
where $\chi(x)$ is the characteristic function of the domain $I$,
$$ \chi(x) := \left\{ \matrix{1, & {\rm if}\ x\in I, \cr 
0, & {\rm otherwise}, \cr} \right. \eqno (1.6)$$
and $R(z,I)$ is the generating function of the integrals over 
$I$ of the correlation functions $R_k(x_1,...,x_k)$, 
$$ \eqalignno{ R(z,I) & := \int F({\bf x})\prod_{j=1}^n\left[1+z\chi(x_j)
\right] dx_j = \sum_{k=0}^n {\rho_k\over k!}z^k, & (1.7) \cr
\rho_k & = \left\{ \matrix{ 1, \hfill & k=0, \hfill \cr 
\int R_k(x_1,...,x_k)\displaystyle\prod_{j=1}^k \chi(x_j)dx_j, &  
{\rm otherwise}. } \right. & (1.8) \cr} $$

The $R(z,I)$  of Eq. (1.7) can be expressed as a determinant 
$$ R(z,I) = \det[G_{ij}]_{i,j=0,...,n-1}, \eqno (1.9)$$
where, using the orthogonality, Eq. (1.3), of polynomials $P_i(x)$ and 
splitting the constant and linear terms in $z$, $G_{ij}$ reads 
$$ G_{ij} = {1\over h_i} \int P_i(x) P_j(x) \exp[-V(x)] [1+z 
\chi(x)] dx = \delta_{ij}+\bar G_{ij}. \eqno (1.10) $$
Finally, $R(z,I)$ can also be written as the Fredholm determinant 
$$ R(z,I) = \prod_{k=1}^n \left[1+\lambda_k(z,I)\right] \eqno (1.11)$$
of the integral equation
$$\int N(x,y) f(y) dy = \lambda f(x), \eqno(1.12)$$
where remarkably the kernel $N(x,y)$  is simply 
$z K(x,y) \chi(y)$ with $K(x,y)$ of Eq. (1.2). 
The $\lambda_i(z,I)$ are the eigenvalues of the above equation and also 
of the matrix $[\bar G_{ij}]$.
\vskip 3mm

These results can be extended to a chain of $p$ complex 
hermitian $n\times n$ matrices. We  
consider the probability density for their elements
$$ \eqalignno{ {\cal F} (A_1,\cdots , A_p)\, & \propto\, 
\exp\left[-{\rm tr}\left\{ {1\over 2} V_1(A_1)+V_2(A_2)+\cdots+V_{p-1}
(A_{p-1})+{1\over 2} V_p(A_p)\right\}\right] & \cr 
& \times\,\exp\left[ {\rm tr} \left\{ c_1A_1A_2+c_2A_2A_3+\cdots
+c_{p-1}A_{p-1}A_p \right\}\right]. & (1.13) \cr}$$
Here $V_j(x)$ are real polynomials of even order with positive coefficients 
of their highest powers and the $c_j$ are real constants such that all the 
integrals which follow converge.  For each $j$ the eigenvalues of the 
matrix $A_j$ are real and will be denoted 
by ${\bf x_j} := \{x_{j1}$, $x_{j2}$, ..., $x_{jn}\}$. The probability 
density for 
the eigenvalues of all the $p$ matrices resulting from Eq. (1.13) is [2-5]
$$ \eqalignno{ & F({\bf x_1};...;{\bf x_p}) 
 = C \exp\left[-\sum_{r=1}^n\left\{ {1\over 2} V_1(x_{1r})+
V_2(x_{2r})+\cdots+V_{p-1}(x_{p-1r})+
{1\over 2} V_p(x_{pr})\right\} \right] & \cr
& \times\left[ \prod_{1\le r<s\le n}(x_{1s}-x_{1r})(x_{ps}-x_{pr})\right] 
\det\left[e^{c_1x_{1r}x_{2s}}\right] 
 \det\left[e^{c_2x_{2r}x_{3s}}\right]\cdots\det\left[e^{c_{p-1}
x_{p-1r}x_{ps}}\right] & \cr & & (1.14) \cr
& = C \left[\prod_{1\le r<s\le n}(x_{1s}-x_{1r})(x_{ps}-x_{pr}) \right]
\left[ \prod_{k=1}^{p-1} \det\left[w_k(x_{kr},x_{k+1s})\right]_{r,s=1,...,n} 
\right], & (1.15) \cr}$$
where
$$w_k(x,y) := \exp\left[-{1\over 2} V_k(x)-{1\over 2} V_{k+1}(y)+
c_k x y \right], \eqno (1.16) $$
and $C$ is a normalisation constant such that the integral of $F$ over all 
the $np$ variables $x_{ir}$ is one.

The correlation function 
$$ \eqalignno{ & R_{k_1,...,k_p}(x_{11},...,x_{1k_1};...;x_{p1},...,x_{pk_p}) 
\cr
& := \int F({\bf x_1};...;{\bf x_p}) \prod_{j=1}^p\left[ {n!\over (n-k_j)!}
\prod_{r_j=k_j+1}^n dx_{jr_j}\right], & (1.17) \cr} $$
was calculated in a previous paper [6] to be an $m\times m$ determinant 
($m=k_1+...+k_p$)
$$ \eqalignno{ & R_{k_1,...,k_p}(x_{11},...,x_{1k_1};...;x_{p1},...,x_{pk_p}) 
& \cr
& = \det \left[K_{ij}\left(x_{ir}, x_{js}\right)\right]_
{i,j=1,...,p; r=1,...,k_i; s=1,...,k_j}. & (1.18) \cr}$$
This is the density of ordered sets of $k_j$ eigenvalues of $A_j$ within 
small intervals around $x_{j1}$, ..., $x_{jk_j}$ for $j=1$, 2, ..., $p$. 

Here we will consider the spacing function $E(k_1,I_1;...;k_p,I_p)$, i.e. 
the probability that the domain $I_j$ contains exactly $k_j$ eigenvalues 
of the matrix $A_j$ for $j=1,...,p$, $0\le k_j\le n$. The domains $I_j$ 
may have overlaps. As in the one matrix case one has evidently  
$$ E(k_1,I_1;...;k_p,I_p) = \left. {1\over k_1!}
\left({\partial\over\partial z_1}\right)^{k_1} ...
{1\over k_p!}\left({\partial\over\partial z_p}\right)^{k_p} 
R(z_1,I_1;...;z_p,I_p)\right\vert_{z_1=...=z_p=-1}, \eqno (1.19) $$
with the generating function 
$$ \eqalignno{ 
R(z_1,I_1;...;z_p,I_p) & = \int F({\bf x_1};...;{\bf x_p}) 
\prod_{j=1}^p \prod_{r=1}^n \left[1+z_j\chi_j(x_{jr})\right] dx_{jr} 
& (1.20) \cr
& = \sum_{k_1=0}^n... \sum_{k_p=0}^n {\rho_{k_1,...,k_p}\over 
k_1!...k_p!} z_1^{k_1}...z_p^{k_p}, & (1.21) \cr}$$
$$\eqalignno{
\rho_{k_1,...,k_p} & = \left\{ \matrix{ 1, \hfill & k_1=...=k_p=0, \hfill \cr 
\displaystyle\prod_{j=1}^p \left[ \int_{I_j}\displaystyle\prod_{r=1}^{k_j} 
dx_{jr} \right] 
R_{k_1,...,k_p}(x_{11},\cdots,x_{1k_1};\cdots;x_{p1},
\cdots, x_{pk_p}), & {\rm otherwise}, \hfill \cr} \right. & \cr 
& & (1.22) \cr} $$
$\chi_j(x)$ being the characteristic function of the domain $I_j$, Eq. (1.6).

The function $R(z_1,I_1;...;z_p,I_p)$ will be expressed as a  
$n\times n$ determinant. It will also be written as a Fredholm 
determinant, the kernel of which will now depend on the variables 
$z_1$, ..., $z_p$ and the domains $I_1$, ..., $I_p$ in a more involved 
way than in the one matrix case. In particular, it does not have the 
remarkable form mentioned after Eq. (1.12).  
\vskip 1cm

\noindent {\bf 2. The generating function $R(z_1,I_1;...;z_p,I_p)$}
\vskip 3mm

\noindent The expression of $F$, Eq. (1.15), contains a product of 
determinants. As the product of differences 
$$\Delta({\bf x_1}) = \prod_{1\le r<s\le n}(x_{1s}-x_{1r}) \eqno (2.1)$$
and $\det[w_1(x_{1r},x_{2s})]$ are completely antisymmetric and other 
factors in the integrand of Eq. (1.20) are completely symmetric in the
variables $x_{11}$, ..., $x_{1n}$, one can replace 
$\det[w_1(x_{1r},x_{2s})]$ under the integral sign in Eq. (1.20) by a 
single term, say the diagonal one, and multiply by $n!$. 
This single term is invariant under a permutation of the variables 
$x_{1r}$ and simultaneously the same permutation on the variables 
$x_{2r}$. Therefore, after 
integration over the $x_{1r}$, $r=1,...,n$, the integrand, excluding 
the factor $\det[w_2(x_{2r},x_{3s})]$, is completely antisymmetric in 
the variables $x_{21}$, ..., $x_{2n}$ and so one can replace the second 
determinant $\det[w_2(x_{2r},x_{3s})]$ by a single term, say the diagonal 
one, and multiply the result by $n!$. In this way, under the integral sign 
one can replace successively each of the $p-1$ determinants $\det[w_k(x_{kr}, 
x_{k+1s})]$ by a single term multiplying the result each time by $n!$
$$ \eqalignno{ 
R(z_1,I_1;...;z_p,I_p) & = (n!)^{p-1} C \int \Delta({\bf x_1})
\Delta({\bf x_p}) 
\left[\prod_{j=1}^{p-1}\prod_{r=1}^n w_j(x_{jr},x_{j+1r})\right] & \cr  
& \hskip 5mm \times \left[\prod_{j=1}^{p}\prod_{r=1}^n 
\Big[1+z_j\chi_j(x_{jr}) \Big] dx_{jr}\right]. & (2.2) \cr} $$

Recall that a polynomial is called monic when the coefficient of the 
highest power is one. Also recall that the product of differences 
$\Delta({\bf x})$ can be written as a determinant 
$$\Delta({\bf x}) = \det[x_i^{j-1}] = \det[P_{j-1}(x_i)] = 
\det[Q_{j-1}(x_i)], \eqno (2.3)$$
where $P_j(x)$ and $Q_j(x)$ are arbitrary monic polynomials of degree $j$. 
As usual, we will choose these polynomials real and bi-orthogonal [6] 
$$\int P_j(x)(w_1*w_2*...*w_{p-1})(x,y)Q_k(y)dx dy = h_j \delta_{jk}, 
\eqno (2.4)$$ 
with the obvious notation
$$ (f*g)(x,y) = \int f(x,\xi)g(\xi,y) d\xi. \eqno (2.5)$$
The normalization constant $C$ is [6],
$$ C = (n!)^{-p} \prod_{i=0}^{n-1} h_i^{-1}. \eqno (2.6) $$
Now expand the determinant as a sum over $n!$ permutations 
$(i) := \left( \matrix{ \scriptstyle 
0, & \cdots, & \scriptstyle n-1 \cr \scriptstyle i_1, & \cdots, 
& \scriptstyle i_n \cr}\right)$, $\pi(i)$ being its sign,
$$\det[P_{s-1}(x_{1r})] = \sum_{(i)} \pi(i) P_{i_1}(x_{11})P_{i_2}(x_{12})
...P_{i_n}(x_{1n}). \eqno (2.7)$$
Doing the same thing for $\det[Q_{s-1}(x_{pr})]$ and integrating over 
all the $np$ variables $x_{jr}$; $j=1,...,p$; $r=1,...,n$ in Eq. (2.2), one 
gets 
$$ \eqalignno{
R(z_1,I_1;...;z_p,I_p) & ={1\over n!} \sum_{(i)}\sum_{(j)} \pi(i) \pi(j) 
G_{i_1j_1}G_{i_2j_2}...G_{i_nj_n} & \cr
& = \det[G_{ij}]_{i,j=0,...,n-1},  & (2.8) \cr}$$
where
$$ G_{ij} = {1\over h_i} \int P_i(x_1)
\left[\prod_{k=1}^{p-1}w_k(x_k,x_{k+1}) \right] Q_j(x_p)
\left[\prod_{k=1}^p [1+z_k\chi_k(x_k)] dx_k\right].
\eqno (2.9)$$

When all the $z_k$ vanish, $G_{ij}$ is equal to $\delta_{ij}$ as a 
consequence of the bi-orthogonality, Eq. (2.4), of the polynomials $P_i(x)$ 
and $Q_i(x)$. Let us define $\bar G_{ij}$ as follows
$$\bar G_{ij} := G_{ij} - \delta_{ij}, \eqno (2.10) $$
so that 
$$ \bar G_{ij} = {1\over h_i} \int P_i(x_1)
\left[\prod_{k=1}^{p-1}w_k(x_k,x_{k+1}) \right] Q_j(x_p)
\left[\prod_{k=1}^p [ 1+z_k\chi_k(x_k)]-1 \right] 
\left[ \prod_{k=1}^p dx_k\right]. \eqno (2.11)$$
Any $n\times n$ determinant is the product of its $n$ eigenvalues and 
therefore one has
$$ R(z_1,I_1;...;z_p,I_p) = \prod_{k=1}^n \left[1+
\lambda_k(z_1,I_1;...;z_p,I_p) 
\right], \eqno (2.12) $$
where the $ \lambda_k(z_1,I_1;...;z_p,I_p) $ are the $n$ roots (not 
necessarily distinct, either real or pairwise complex conjugates, 
since $\bar G_{ij}$ is real) of the algebraic equation in $\lambda$ 
$$ \det[\bar G_{ij}-\lambda \delta_{ij}] = 0. \eqno (2.13)$$
One can always write a Fredholm integral equation with a separable 
kernel whose eigenvalues are identical to these (cf. reference [7] for 
the case of $p=2$ matrices). Indeed, for any eigenvalue $\lambda$ the 
system of linear equations 
$$ \sum_{j=0}^{n-1} \bar G_{ij} \xi_j = \lambda \xi_i \eqno (2.14) $$
has at least one solution $\xi_i$, $i=0$, ..., $n-1$, not all zero.
Multiplying both sides of the above equation by $Q_i(x)$, summing over $i$ 
and using Eq. (2.11) gives the Fredholm equation 
$$ \int N(x,x_p) f(x_p) dx_p = \lambda f(x), \eqno (2.15) $$
where 
$$ \eqalignno{ f(x) & := \sum_{i=0} ^{n-1} \xi_i Q_i(x), & (2.16) \cr
N(x,x_p) & := \sum_{i=0}^{n-1} {1\over h_i} Q_i(x) \int P_i(x_1)
\left[\prod_{k=1}^{p-1}w_k(x_k,x_{k+1}) \right] 
\left[\prod_{k=1}^p [ 1+z_k\chi_k(x_k)]-1 \right] 
\left[\prod_{k=1}^{p-1}
dx_k \right]. & \cr & & (2.17) \cr} $$
Hence if $\lambda$ is an eigenvalue of the matrix $[\bar G_{ij}]$, 
it is also an eigenvalue of the integral equation (2.15). Conversely, 
since the kernel $N(x,x_p)$ is a sum of separable ones and since 
the polynomials $Q_i(x)$ for $i=0$, ..., $n-1$ are linearly 
independent, if $\lambda$ and $f(x)$ are, respectively an eigenvalue 
and an eigenfunction of this integral equation, then $f(x)$ is 
necessarily of the form 
$$ f(x) = \sum_{i=0} ^{n-1} \xi_{i} Q_i(x), \eqno (2.18) $$
and the $\xi_{i}$, $i=0$, ..., $n-1$, not all zero, satisfy Eq. (2.14). 
Therefore $\lambda$ is a root of Eq. (2.13).

When one considers 
the eigenvalues of a single matrix anywhere in the chain, disregarding 
those of the other matrices, everything works as if one is dealing with 
the one matrix case and formulas (1.2), (1.5), (1.7) and (1.11) are valid 
with minor replacements. Similarly, when one considers properties of 
the eigenvalues of $k$ ($1\le k\le p$) matrices situated anywhere in 
the chain, not necessarily consecutive, everything works as if one is 
dealing with a chain of only $k$ matrices; the presence of other 
matrices modifying only the couplings. 

To say something more about the general case seems difficult. 

When $V_j(x)=a_jx^2$, 
$j=1$, ..., $p$, then the polynomials $P_j(x)$ and $Q_j(x)$ are 
Hermite polynomials $P_j(x)=H_j(\alpha x)$, $Q_j(x)=H_j(\beta x)$, 
the constants $\alpha$ and $\beta$ depending on the parameters $a_j$ and 
the couplings $c_j$. In this particular case the calculation can be 
pushed to the end (see appendix). 
\vskip 1cm

\noindent {\bf Appendix} 
\vskip 3mm

\noindent For $V_j(x) = a_j x^2$, $j=1$, ..., $p$, setting 
$$ W_{a,b,c}(x,y) := \exp \left(-{1\over 2}ax^2-{1\over 2}by^2+cxy \right), 
\eqno (A.1)$$
one gets according to Eq. (2.5) the multiplication law 
$$\left(W_{a,b,c}*W_{a',b',c'}\right)(x,y) 
= \left({2\pi\over b+a'}\right)^{1/2} W_{a'',b'',c''}(x,y), \eqno (A.2)$$
where
$$a''=a-{c^2\over b+a'},  \hskip 5mm b''=b'-{c'^2\over b+a'},   
\hskip 5mm c''={cc'\over b+a'}. \eqno (A.3)$$
From Eq. (1.16), $w_k(x,y)=W_{a_k,a_{k+1},c_k}(x,y)$ and a repeated use of 
the above multiplication law yields 
$$W(x,y) := (w_1*w_2*\cdots *w_{p-1})(x,y) = d\times 
W_{a,b,c}(x,y), \eqno (A.4)$$ 
where $a$, $b$, $c$ and $d$ are constants depending on the parameters 
$a_1$, ..., $a_p$ and $c_1$, ..., $c_{p-1}$.

The orthogonality relation (2.4) of the polynomials $P_j(x)$ and $Q_j(x)$ 
takes the form 
$$\int P_j(x) W(x,y) Q_k(y) dx dy = h_j \delta_{jk}, \eqno (A.5)$$
namely the same relation as in the two matrix case with the 
weight $W(x,y)$, an exponential of a quadratic form in $x$ and $y$. 
It follows that $P_j(x)$ and $Q_j(x)$ are Hermite polynomials of 
$x$ times a constant 
$$ \eqalignno{ & P_j(x) = H_j(\alpha x), \hskip 1 cm  \alpha := 
\left( {ab-c^2\over 2b}\right)^{1/2}; & (A.6) \cr
& Q_j(x) = H_j(\beta x), \hskip 1 cm  \beta :=
\left( {ab-c^2\over 2a}\right)^{1/2}; & (A.7) \cr 
& h_j = {2\pi\over (ab-c^2)^{1/2}}
\left( {c\over \sqrt{ab} }\right)^j 2^j j!\ d. & (A.8) \cr}$$
The eigenvalue density of the matrix $A_1$, for example, ignoring the 
eigenvalues of other matrices, is from Eq. (1.18) 
$$ \eqalignno{ R_1(x) & = K_{11}(x,x) 
= \sum_{j=0}^{n-1} {1\over h_j} P_j(x) \int W(x,y) Q_j(y) dy & \cr
& = d\left({2\pi\over b}\right)^{1/2} e^{-\alpha^2 x^2}\sum_{j=0}^{n-1} 
{1\over h_j} \left({c\over \sqrt{ab}}\right)^j H_j^2(\alpha x) & \cr
& = {\alpha\over \sqrt\pi} e^{-\alpha^2x^2} 
\sum_{j=0}^{n-1} {H_j^2(\alpha x)\over 2^j j!}, & (A.9) \cr}$$
which in the large $n$ limit is a semi-circle of radius $\sqrt {2n}/\alpha$.
Thus in this particular case of coupled matrices one recovers Wigner's 
``semi-circle law" for the eigenvalues of a single matrix.

The kernel of the integral equation (2.15) is given by Eq. (2.17) with 
$P_j(x)$, $Q_j(x)$ and $h_j$ as given above. To get further, one has to 
take explicitly the domains $I_j$ into account. 

\vskip 1cm

\noindent {\bf References}
\vskip 3mm

\item{1.} See for example, M.L. Mehta, {\it Random Matrices}, Academic Press,
San Diego, CA, U.S.A., 1991, Chapter 3. There only the case $V(x)=x^2$ is 
considered, but the same method applies when $V(x)$ is any real polynomial of 
even order. 

\item{2.} See for example, reference 1, appendix A.5.

\item{3.} C. Itzykson and J.-B. Zuber, The planar approximation II, J. Math. 
Phys. 21 (1980) 411-421. 

\item{4.} M.L. Mehta, A method of integration 
over matrix variables, Comm. Math. Phys. 79 (1981) 327-340.

\item{5.} M.L. Mehta, {\it Matrix Theory}, Les Editions de Physique, 
Les Ulis, France, 1989, Chapter 13.2.

\item{6.} B. Eynard and M.L. Mehta, Matrices coupled in a chain. I. 
Eigenvalue correlations. Saclay preprint T97/112 
(submitted to J. Phys. A: Maths. and General).

\item{7.} M.L. Mehta and P. Shukla, Two coupled matrices: eigenvalue 
correlations and spacing functions, J. Phys. A: Math Gen. 27 (1994) 
7793-7803.

\end